\documentclass[12pt,sort&compress]{elsarticle}

\usepackage{graphicx}
\usepackage{amssymb}
\usepackage{amsthm}
\usepackage[nodots]{numcompress}
\usepackage{epsfig}
\usepackage{epstopdf}
\usepackage{subfig}
\usepackage{amsmath}
\usepackage{latexsym}
\usepackage{amsbsy}
\usepackage{array}
\usepackage{setspace}
\usepackage{bm}
\usepackage{textcomp}

\journal{--}
\begin{document}

\begin{frontmatter}

\title{A Physical Review on Currency}

\author[sjtu,rizt]{Ran ˜Huang}
\ead{ranhuang@sjtu.edu.cn}
\address[sjtu]{School of Life Science and Biotechnology, Shanghai Jiao Tong University, Shanghai 200240, China}
\address[rizt]{Department of Materials Science and Engineering, Research Institute of Zhejiang University-Taizhou, Taizhou, Zhejiang 318000, China}

\begin{abstract}
A theoretical self-sustainable economic model is established based on the fundamental factors of production, consumption, reservation and reinvestment, where currency is set as a unconditional credit symbol serving as transaction equivalent and stock means. Principle properties of currency are explored in this ideal economic system. Physical analysis reveals some facts that were not addressed by traditional monetary theory, and several basic principles of ideal currency are concluded: 1. The saving-replacement is a more primary function of currency than the transaction equivalents; 2. The ideal efficiency of currency corresponds to the least practical value; 3. The contradiction between constant face value of currency and depreciable goods leads to intrinsic inflation.

\end{abstract}


\end{frontmatter}

\section{Introduction}
Traditional monetary theory well defined the currency, or to say, money, with the functions of exchange medium, measure of value, standard of deferred payment, and store of value\cite{wiki}. Putting aside how the currency evolved and developed in history\cite{his}, nowadays these functions are roundly treated as the basic properties of currency, and they are believed to be originated from the identity of exchange medium. However, by starting from the first principle, a pure theoretical analysis on simple physical model may reveal that, regardless of the actual appearance sequence in economic history, the fundamentals of currency may not arise from the function of exchange.

In this work, a theoretical economic model is interpreted with rational participants and their abstract economic activities such as production, consumption, reservation, reinvestment and transactions. With the restrictive conditions of individual sustainability and maximum systematic welfare, the functions and operations of an ideal currency has been investigated and defined by several basic principles. Beside the results are found to be consistent with classical monetary theories, further investigation also provides insights into some mysterious and controversial issues of currency, e.g. the monetary standard, money supply and inflation.

\section{Model}

From a fundamental view, an economic system can be understood as an assemble of individuals with subjective initiative doing operations for self-sustain by resourcing and processing useful matters from environment. We denote the behavior individual in the economic system by the term ``participant" ($ P $), the ``useful matters" by the term ``goods" ($ G $), and understand the ``resourcing and processing" as production in general. Other than a non-interested hunter-gatherer society, the goods is set to be obtained from the investment of identical $ G $, and the production multiplies initial $ G $s to be times of replicas as products, which are to be consumed, reserved for emergence and reinvestment for the following productions.    
\subsection{The simplest sustainable model of one body and one good}\label{sSCR}

\begin{figure}
	\centering{
		\includegraphics[width=0.5\textwidth]{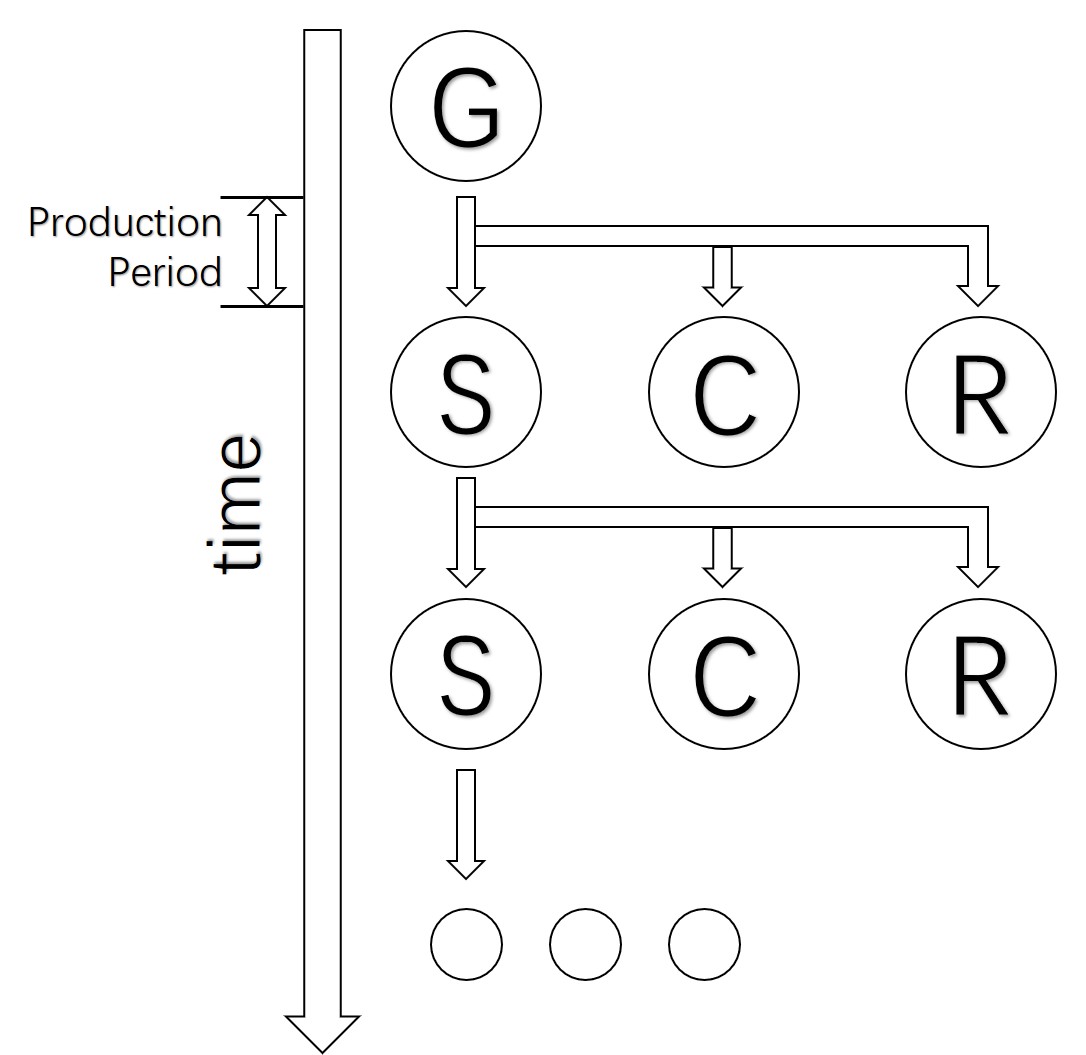}
		\caption{The demonstration of self-sustainable one-body SCR model.}
		\label{fig1}
	}
\end{figure}

For the simplest case, suppose there is only one participant, who lives indispensably onto one type of goods. In a so-called ``production period", the participant invests an initial unit of G to produce three identical replicas, and suppose that one $ P $ needs to consume one unit of $ G $ in one production period, then the products can be assigned into three categories, one is kept as the raw materials for investment in the next period ($ S $ as ``seed"), one is the normal consumption ($ C $), and the other is stocked as reservation ($R$) for any general risk, while it is assumed that the mechanism of reservation is necessary for sustainability in the long run. This setup is referred as ``SCR" model in the following discussion. The model is illustrated in Fig.\ref{fig1}.

Although this model is designed to describe a general economic production process, for an easy understanding, we may comprehend this system as a primitive agricultural economics, that one $ P $ plants a number of seeds and harvests three times amount of grains, then a third of them are consumed, a third are kept as seeds for next planting, and the other one third are reserved for emergencies. Additionally, it is reasonable to set the expiration of the goods to be one period, i.e. the stock $ R $ depreciates to zero right at the time point where new goods are produced. The setup and parameters are for convenience in the qualitative analysis and might be too ideal to describe the reality, but the principles and underlying mechanism should be acceptable.

The production increment is with the resource from environment and/or the participant's input (e.g. labor force), in this work we will ignore this part and simply understand it as a spontaneous factor for the participant's self-sustainability.

Note at this stage, currency functioned as a transaction equivalent is unnecessary, however we will show that it can execute a more primary function even in the single-body system, contrast to the traditional sense that the first-principle role of money is transaction equivalent.

\subsection{The SCR model of multi-participants and multi-goods}\label{mSCR}

\begin{figure}
	\centering{
		\includegraphics[width=0.8\textwidth]{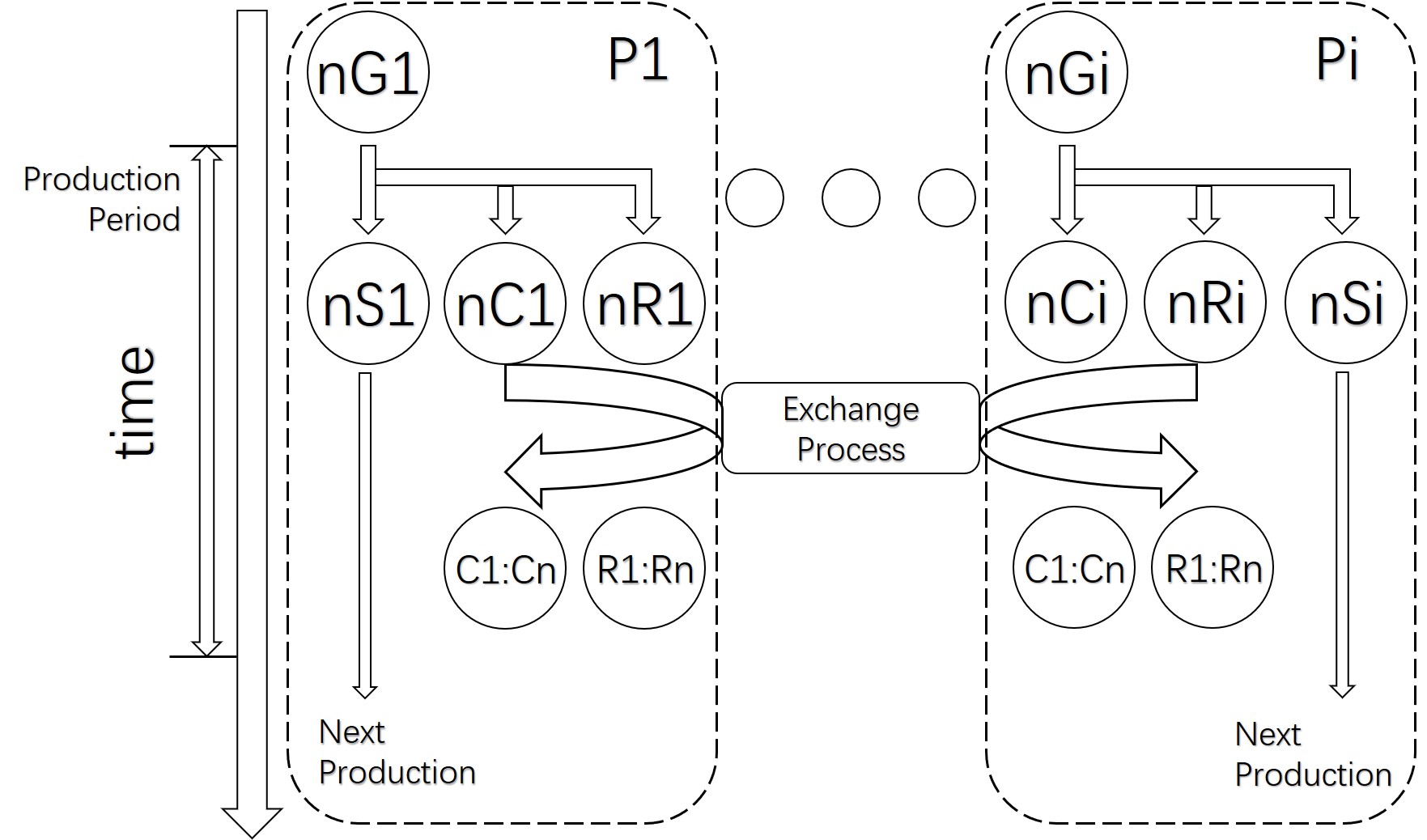}
		\caption{The demonstration of self-sustainable many-body SCR model. Note in text we have only described the exchanges of goods for consumption part $C_i$, for a better illustration in this figure the exchanges of $R_i$ are also included, which however is a trivial operation.}
		\label{fig2}
	}
\end{figure}

Now assume there are $ n $ types of indispensable goods $ G_1 $, $ G_2 $, ... and $ G_n $ produced by $ n $ participants, each $P$ only produces one type of $G$, i.e. $P_i$ with $G_i$, but each $P$ must consume exactly one each of these $G$s to survive during every production period. The productivity of all $G$s are still set to be $ 1:3 $, then it is reasonable to treat all $G$s to be equally valued as of one ``unit" by all participants. After each harvest, every $P_i$ holds the stock of $ nC_i $ to transact with others, and eventually everyone will obtain one of $ C_1 $, $ C_2 $, ... and $ C_n $ to fulfill its consumptions. The shortage of any $C_i$ will drive a participant to do the exchange until its necessary consumption of all types of $G$s is fulfilled\footnote{Geometrically our model is hyper-dimensional, i.e. all the participants have even and immediate accessibility to each other. The model can be extended to be more complicated economics if we map the system on finite dimensions, then there will be accessibility barriers, goods circulation delay and the consequent arbitrage opportunities, where the business arises. }. The operation of many-body SCR is shown in Fig.\ref{fig2}.

It is clear that during each period, the system has a driving force to achieve the even distribution of all $C$s to all $P$s. We introduce the dispersity $ D $ in the fashion of variance, to measure the distribution/concentration degree of $C$s in many-body SCR:

\begin{equation}
D=\dfrac{\Sigma(X_i-Xe_i)^2}{n-1} \label{dispersity}
\end{equation}
where $ X_i $ is the quantity of some goods holding on the $ i $th participant, $ Xe_i $ is the quantity expected to be held by $ P_i $, which is just one in the present case. Obviously, $D$ has its maximum value at the beginning of exchange, and approach to 0 along with transaction process. And the concentration degree, which can be defined as 

\begin{align*}
Con\%=\dfrac{D}{D_{max}}\times100\%,
\end{align*}
goes from 100\% to 0. Although these quantities are lack of interest for the present simple model, they will be useful to track the distributions of goods and currency (yet introduced) in the following sections. 

\begin{figure}
	\centering{
		\includegraphics[width=0.8\textwidth]{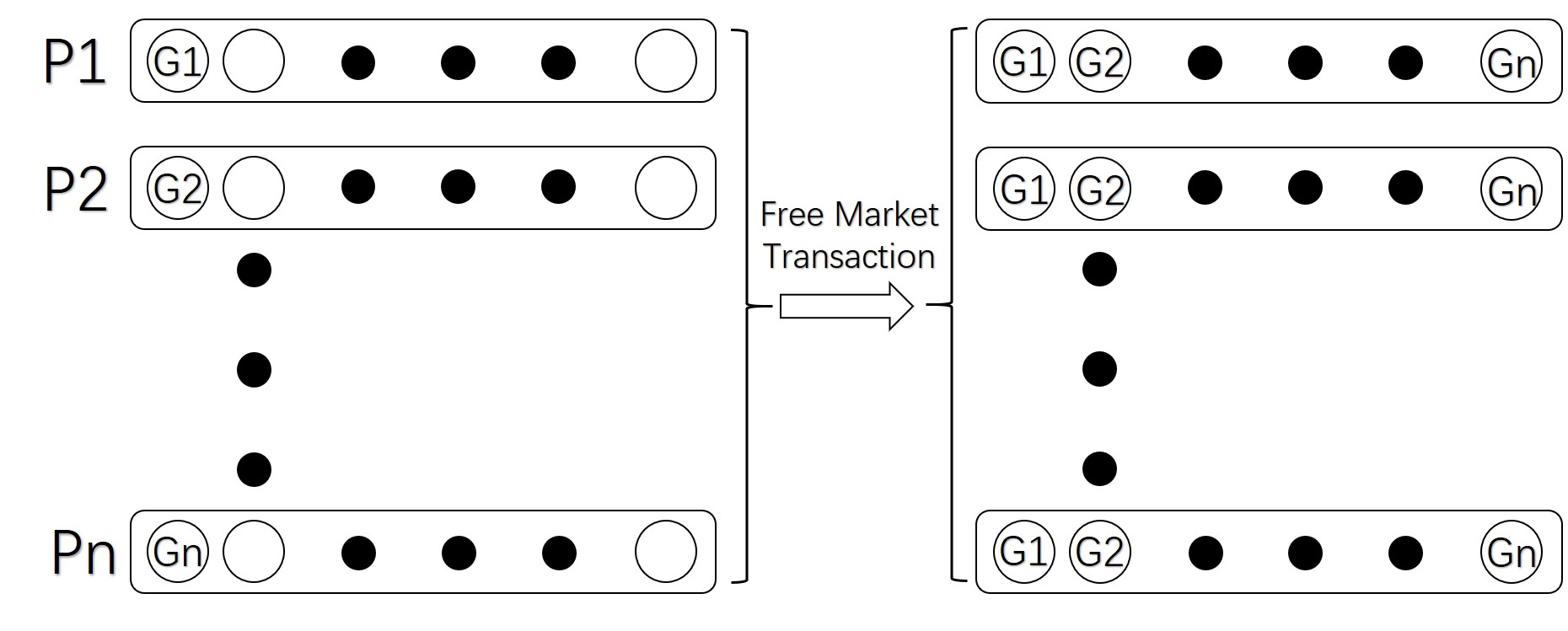}
		\caption{The free market transaction to achieve the expected goods distribution in many-body SCR model.}
		\label{fig3}
	}
\end{figure}

At this stage, currency as a transaction equivalent is still unnecessary, either with or without an exchange medium, all $P$s have to deal with each other at least once (shown in Fig.\ref{fig3}), a currency cannot reduct the total number of transactions $ n(n-1)/2 $ and enhance any efficiency, therefore the barter trade is sufficient for the transaction need.

\section{Presence of currency}

First of all, we define the ex ante law of an ideally functional currency: it is an artificial symbol holding unconditional creditability on a constant face value to measure the utility of real goods, and this face value is fully appreciated by all the participants that it can homogeneously transact to any transferable goods without any restriction in the system.

In this work, the economics is assumed to be ideal with an imaginary heat reservoir\footnote{The ``heat reservoir" is a concept in thermodynamics, which provides unconditional energy resource to the interested thermophysical system, we borrow this idea here to construct our model to be a physically ideal system.} to provide authoritative currency. The money issue mechanism and the cost of currency system are neglected at present. The properties in reality such as the cost and internal value of the currency will be discussed later.

\subsection{The saving-replacement function} \label{sav}

\begin{figure}
	\centering{
		\includegraphics[width=0.7\textwidth]{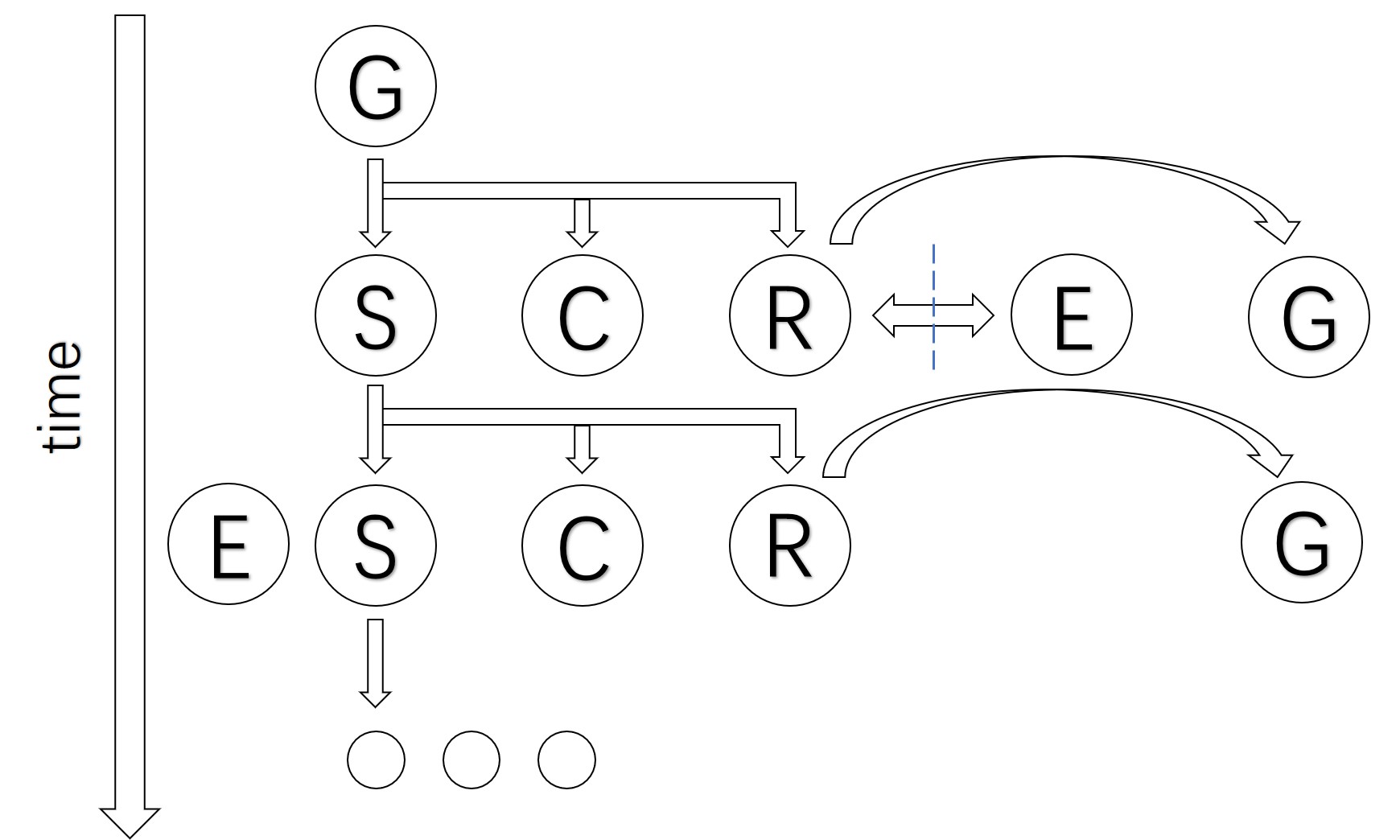}
		\caption{The demonstration of saving-replacement with exchange symbol in one-body SCR model.}
		\label{fig4}
	}
\end{figure}

With the assumptions above, figure \ref{fig4} demonstrates the effect of introducing currency into a single-body SCR model. After the first production period, in which the initial unit $ G $ is produced to be one unit of $ S $, $ C $ and $ R $. While the $ S $ and $ C $ remains respective function shown in Fig.\ref{fig1}, one unit of credible equivalent $ E $ from reservoir can substitute the $ R $ unit out to the environment. In this way, one unit of $G$ which was supposed to be depreciated in stock, can now be utilized as a general goods to be reinvested as $ S $ or consumed as $ C $. Furthermore, since the unconditional credit on the constant value of $ E $, the stock of $ E $ can be inherited into followings periods as constant reservations, i.e. without additional currency input, the $ P $ can always provide one extra goods for general purpose during each production period thereafter, until emergency happens and it has to exchange back $ G $ from the environment. This is defined as the ``Saving-Replacement Function" of currency (SRF).

The extra $ G $ presented in each period may have three possible outcomes: unnecessary consumption (luxury), being abandoned, or extra reservation. And since we have already revealed the advantage of constant value of $ E $, taking the assumptions that $P$ in our model is rational to avoid the first two options, it can be safely assumed that a participant will continue to reserve this extra $ G $ in the form of currency if there are available ones in the system. Hence, in each period, one $P$ will demand one unit of $E$ and a $G$ is continuously substituted out to the environment.

Regardless of the distribution mechanism, this extra $ G $ can be assigned to another $P$ to start an identical production process as a ``new $S$", and again with currency, extra $G$s are recursively passed to the environment, then the entire process makes economic expansion possible. Or in another way, one $P$'s extra product can exactly afford another participant's periodic consumption, who chooses not to produce. Whatever this non-producing participant do, to cause a reason that the system allows its existence, we may define this type of participant as ``non-productivity $ P $" ($nP$) in this work.

With the current setup of the model, for a finite system without goods waste, we have

\begin{equation}
P=nP. \label{np}
\end{equation}

We need to further clarify the physical driving force for the $ P $'s willingness to substitute the $ R $ to be $ E $. Two reasons can be illustrated here: Firstly, the constant value property enables $ P $ to opt out the disadvantage of reservation depreciation for sequential production periods; Secondly, since the significant advantage of saving-replacement, either to extend the economy with additional $P$s or to support $nP$s, the system will encourage this exchange with incentives, for instance the constant value commitment itself, or additional interests. Nevertheless, by just understanding how the mechanism works, to keep our model simple we will not incorporate interest in this work, and the intrinsic driving force of saving-replacement to benefit both the individual $P$s and system can still be validated.

\subsection{Model with Central Agency}\label{central}

\begin{figure}
	\centering{
		\includegraphics[width=0.8\textwidth]{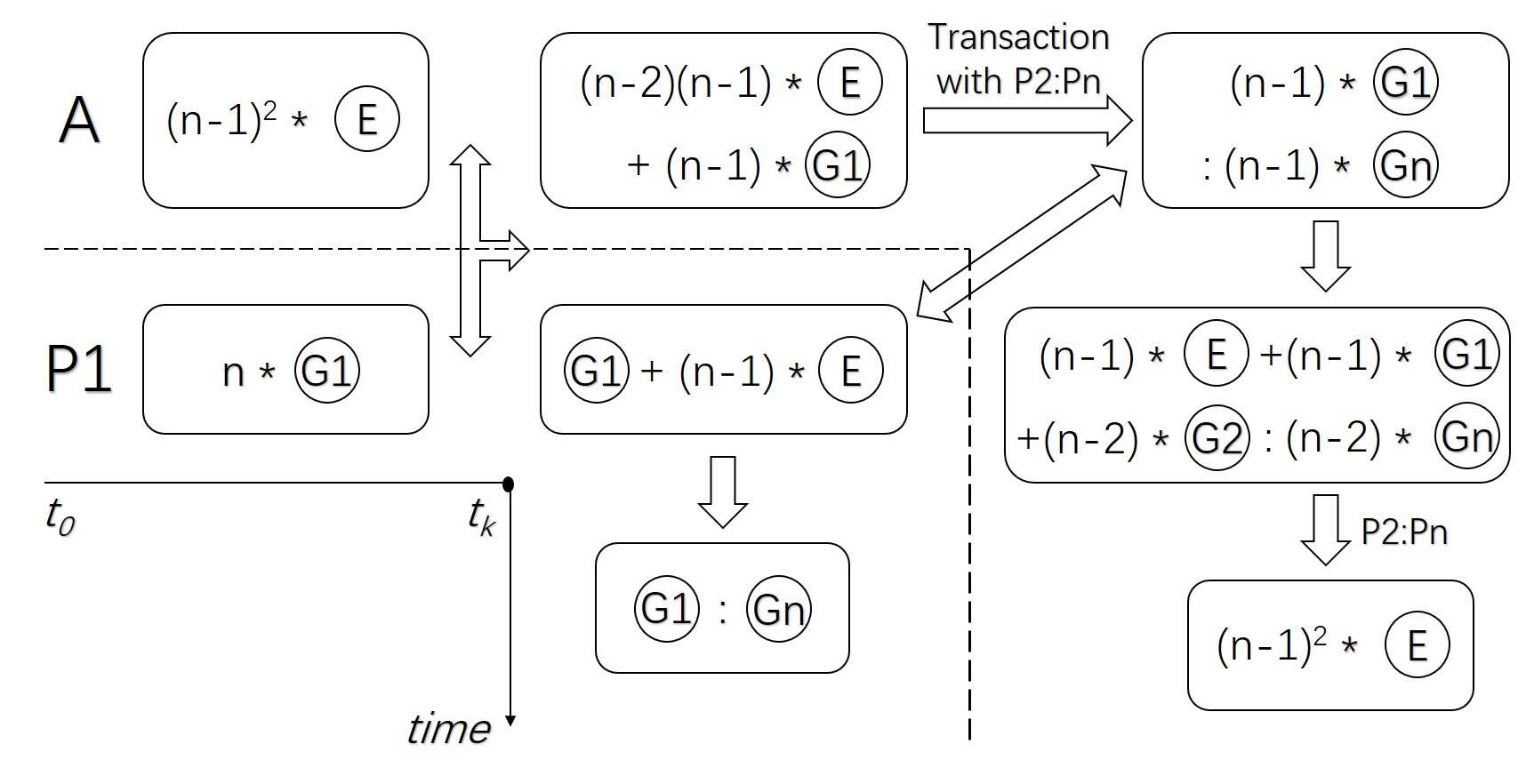}
		\caption{The most effective transaction process to achieve the expected goods distribution in many-body SCR model with exchange agency (A).}
		\label{fig5}
	}
\end{figure}

As analyzed in section \ref{mSCR}, the currency functions as exchange equivalent is unnecessary even in many-body SCR.   Either with fiat money or a random $ G_i $ serving as barter token, the $ n(n-1)/2 $ times transactions are irreducible. However, with an agency which can centralize the transactions, currency can then execute its advantage of exchange medium to enhance the economic efficiency. Figure \ref{fig5} demonstrates the shortest process of goods transaction/distribution with central agency and currency: At stage $ t=t_0 $,  $P_1$ holds $ n$ transferable $ G_1 $s, which is the $ C $ portion, and the agency prepares $(n-1)^2$ exchanges $E$ (for convenience we also set the central agency having the right of issuing). In the first transaction, $P_1$ keeps one of $G_1$ for self-consumption and exchanges the other $n-1$ units to the agency, similar transactions process between the sequential $ P_2 $ to $ P_n $ and agency, until $ t=t_k $ the currency stock in agency has been exhausted and every $P_i$ holds one $G_i$ and $(n-1)E$. Then, again sequentially $ P_1 $ to $ P_n $ transacts with agency reversely to obtain one of $G_1$ to $G_n$ excluding the $G_i$ for $P_i$, with $(n-1)E$. Overall this process only involves $ 2n $ transactions\footnote{Note that transactions among $P$s are allowed but this will not reduce the minimum, every $P$ must visit the agency at least twice for a fulfillment of $G$s.}, which is the minimum and far shorter than the $ n(n-1)/2 $ in the case of free market in many-body system with barter trade for a large $n$.

\begin{figure}
	\centering{
		\subfloat[]{
			\includegraphics[width=0.45\textwidth]{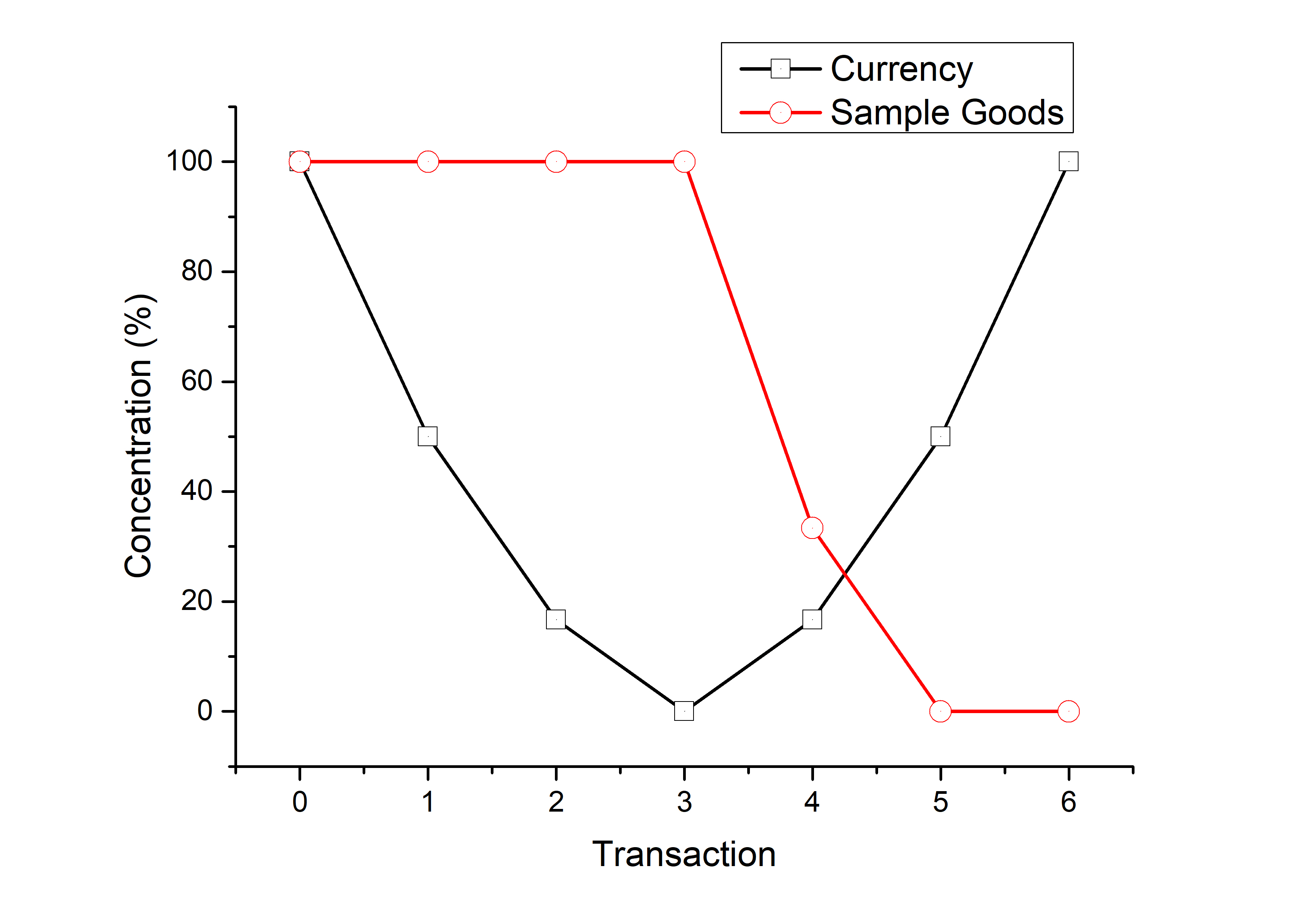}
		}
		\subfloat[]{
			\includegraphics[width=0.45\textwidth]{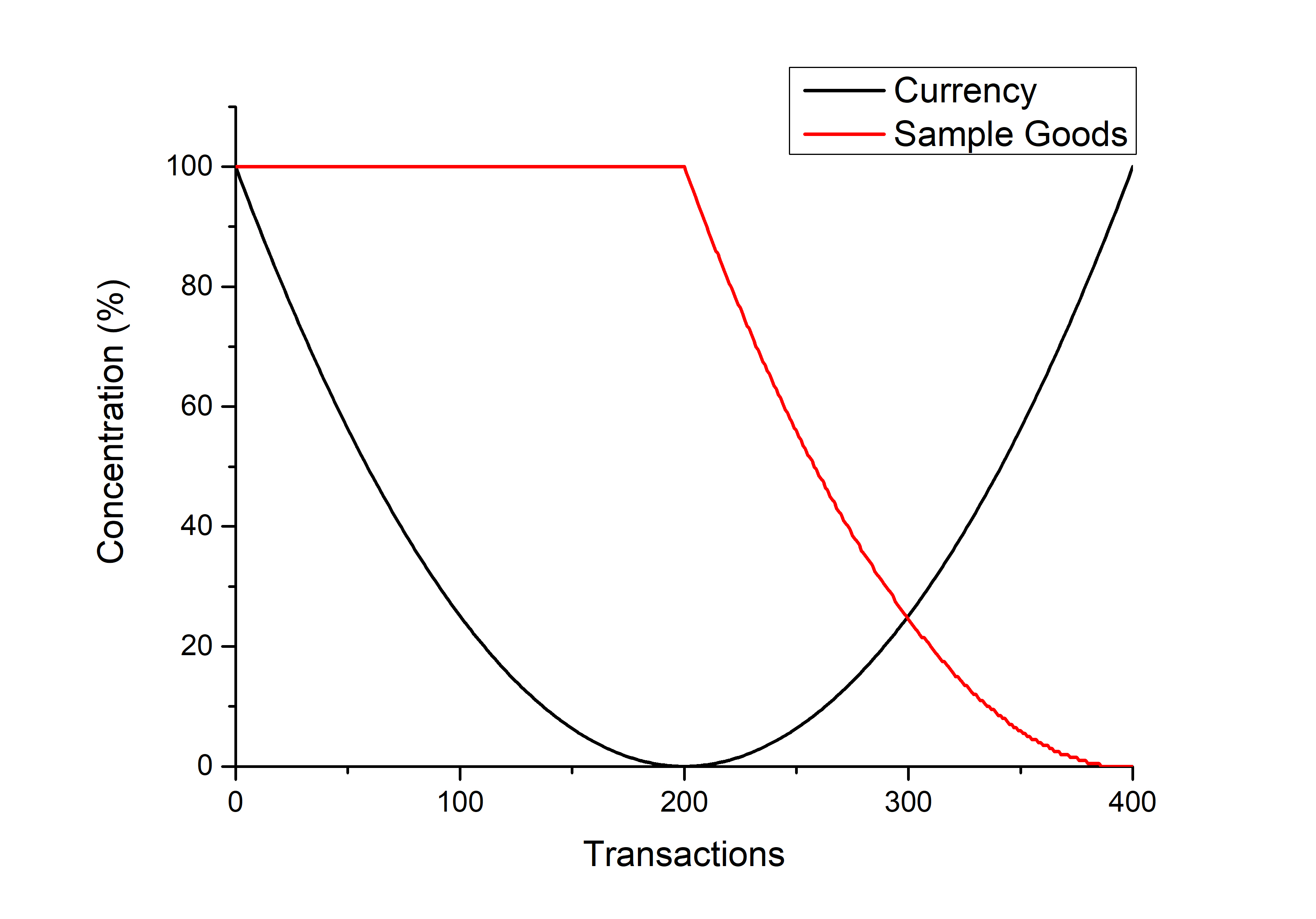}
		}
		\caption{The distributions correlation between a random goods and currency in the shortest transaction process to achieve the expected goods distribution in many-body SCR model with a central agency: (a)$ n=3 $ (b)$ n=200 $. }
		\label{fig6}
	}
\end{figure}

Recall eqn.\ref{dispersity}, by assuming the currency are ``expected" to be held by agency and the $nG$s are ``expected" to be evenly distributed, the concentration of currency and a random $G$ in the process are calculated and presented in fig.\ref{fig6}.

We can see that at the stage $t_0<t<t_k$, the concentration of $E$ drops to zero as the agency stocks are evenly passed to $P$s, and the concentration of a random $G_i$ remains constant; after $t_k$, $G_i$ starts its distribution along with the re-cumulation of $E$ back to the agency, a qualitative description of this process can be given by
\begin{equation}
A_E\varpropto A_G^{-1}, t>t_k \label{Sec}
\end{equation}

where $A$ stands for ``aggregation". The detailed numerical relationship is trivial since figures \ref{fig5} only describes the ideal process, while the possible distribution paths in reality can be much more complicated, nevertheless we can validate the general conclusion that in $t>t_k$ the dispersities of goods and currency are negatively correlated.


It is obvious that this mechanism works in its full efficiency when the central agency issues no less than $ (n-1)^2 $ currency to exchange all the transferable goods at the stage $ t_k $, in this way, it quantifies the advantage of currency acting as transaction equivalent, while less then a certain amount of money will increase the number of necessary transactions to achieve an expected goods distribution (which is even in this case).

One way to another, more money than $ (n-1)^2 $ will either be useless or simply affect the nominal price. Given that the number of necessary transactions to achieve the expected goods distribution is linear to the number of transferable goods, regardless of the supply amount and the initial allocation of currency, we may rewrite the Fisher's theory\cite{minsh} in the form of 
\begin{equation}
T=MV/P=\gamma C_T, \label{Fisher}
\end{equation}
where $T$ is the number of transactions, $M$, $V$ and $P$ stand for their original definitions, $C_T$ is the number of transferable goods for consumption, $\gamma$ is some coefficient relates $T$ and $C_T$, which is $ 2 $ in the present case. From eqn.\ref{Fisher} we can see that the amount of money and price are variables reacting with themselves, and the quantity of transactions is  exclusively related to the number of transferable goods, both are independent to money amount or nominal price. This accords well to Friedman's famous quote:``\textit{Inflation is always and everywhere a monetary phenomenon... }"\cite{Friedman}. 

For the case of more money than $ (n-1)^2 $ issued for $n^2G$s, the increase of nominal price is solo due to the ratio of $ G $ and $ E $ in economic circulation, thus we define it as the ``circulation inflation", or type I inflation. We will figure out some more complex mechanism of inflation in a few sections.  

\subsection{The efficiency of currency}\label{eff}

So far we have demonstrated the advantages of currency in two categories: the saving-replacement function and exchange medium; we have also mentioned the term ``efficiency" in both sections, and it is necessary to distinguish the different efficiency under the two contexts, that one is about the saving-replacement, and maybe measured by how many $nP$s can be supported by $P$s; the other is about the minimal number of transactions to achieve the expected goods distribution. The latter is a relatively simple mathematic measure and has been fully discussed in the previous section. The former one is more of interests and will be detailed in this section. Now and thereafter in this paper, the term ``efficiency" will exclusively refer to the enhancement on economics by SRF.

The eqn.\ref{np} is concluded with the hypothesis that actual value of $E$ is zero, and the $R$ substituted out in fig.\ref{fig4} are full of a unit value and functions 100\% of a $G$. However in many (if not all) realities, this is not true and $E$ contains more or less value that can be measured by a portion of $G$. This actual value of $E$ could be due to the intentional minted value, the cost of materials it is loaded on, or the cost of operating the currency system. We can quantify the efficiency by the following equation 

\begin{equation}
\textit{Eff}=\dfrac{V_F-V_C}{V_F} \leq 1, \label{eq3}
\end{equation}

where $ V_F $ is the face value, and $V_C$ is the actual cost of currency. By this expression, \textit{Eff} indicates the possible partition of $nP$ supported by one $P$, e.g. the boundary condition of $ V_F= V_C$ provides null efficiency and none of $nP$ can be ever supported, while $V_C=0$ leads to eqn.\ref{np}.

Therefore, in the context of SRF, the utility of currency is the credit that endorses its function, i.e. all participants hold unconditional faith that the imaginary symbol can exchange real goods under any circumstance anytime, instead of the real practical utility of the currency contains. Only in the situation of lacking faithful endorsement that $ P $s may require the intrinsic value for the currency's self-endorsement, however this leaves a dilemma that if a currency unit only contains partial value of its face, the credit is not fully self-endorsed, or if it contains the full value of face, then the replacement efficiency is lost at all\footnote{A further phenomenical discussion maybe noted here: Precious metals happen to have none industrial utility appearing to pre-industrialized economy when they were executing the function of currency, if we assume the utility of decoration and enhancement on human psychological satisfaction do not contribute to economical productivity. And though this factor does not account any scientific judgment, the industrial value of gold just appears in the era of modern industrialization, especially the electronics industry, and coincidently, on a large historical time scale, it happens the abandon of gold standard. Some type of cryptocurrency might be strayed into a trap when it is claimed to hold actual worth to be ``valid currency" by the huge consumption of electrical power for `mining'.}.

\subsection{Amount of currency}\label{amount}

With the clarification of currency functioned in saving-replacement and transaction equivalent, the amount of money supply should also be categorized into two meanings: exchange symbol supplied for transaction (denoted as $E_T$) and for the reservation ($E_R$). For the transaction purpose, we have already reformed the Fisher's theory as eqn.\ref{Fisher} in section \ref{central}, that the demand linearly relates to the number of transferable goods produced in one production period, to achieve the minimal number of transactions. Figure \ref{fig6} shows that the $E_T$ re-accumulates back to central agency and can be reused in the next period, therefore, except the natural production growth, $E_T$ does not need to be issued more, while in section.\ref{sav} we concluded that $E_R$ is always demanded in every period\footnote{This may in another way imply that the saving-replacement is a more prior function than the equivalent.}.

There is no financial mechanism of credit/debit in our model, however the concept ``interest" can still be in consideration. If we take the interest as some incentives for participants to hold currency instead of reservation goods, while as discussed in section \ref{sav}, that participants systematically prefer stocking the currency with constant value against to the depreciable goods, the interest rate $ i $ is thereafter the depreciation difference between constant currency and rotting-off goods, i.e. $i$ is the depreciation rate of goods in our model. Notwithstanding the exact value of $i$ is not important (which presently is 100\% per $ t $), a positive $i$ ensures that participants will inelastically exhaust the $ R $ in every period $ t $ to be replaced by currency. In this way, the money demanded for permanent reservation stock, or in another word, the liquidity ``trapped" in participants is

\begin{equation}
L(i,R\cdot t)=\epsilon(i)(\alpha\cdot R\cdot t +\beta),  \label{Keynes}
\end{equation}

where $ \epsilon (i) $ is the step function:

\begin{align*}
\epsilon (i)= 
\begin{cases} 
0, & i\leq0\\ 
1, & i>0 
\end{cases}
\end{align*}

to indicate that an sufficiently small positive $i$ ensures the existence of $ L $. The general linear parameters $ \alpha$ and $ \beta $ are $ 1 $ and $ 0 $ in the present model. 

In summary, the amount of money demanded ($E_d$) to achieve both the requirements in reservation and transaction is 
\begin{equation}
E_d=E_R+E_T=L(i,R\cdot t)+\gamma C_T.
\end{equation}

For a constant economy, over time the $\gamma C_T $ will eventually be neglectable to $ L $, taking $\gamma C\approx 0 $ we have $ P=M/R= L(i,R\cdot t) / R $, that with a constant production to yield $ R $, the relative price increases over time, which implies an inevitable inflation. Recall the discussion on Type I inflation in section \ref{central}, it is clear that there are two identical mechanisms of inflation, while the Type I is more like a numerical manipulation of $E_T/C_T$, the latter appears to reveal an intrinsic contradiction of currency itself. We define it to be ``reservation inflation", or ``type II". If the currency is not designed with a depreciation synchronized to the rotting-off velocity of real goods, this type of inflation may not be avoidable.

\section{Conclusion}
From the above analysis we can summarize the following principles of ideal currency in a general physical economic system:

1. An ideally functional currency should hold unconditional constant credit being homogeneously equivalent to the value of all goods circulating in the economic system, both in the means of reservation and transaction. 
	
	2. The saving-replacement is the priori function of currency to enhance the economic efficiency. The net realistic worth of currency is negatively correlated to the efficiency enhancement. 
	
	3. central agency with currency issue right and goods exchange function can significantly reduct the necessary number of transactions to achieve the expected goods distribution, which is associated with the re-cumulation of currency.
	
	4. In the central agency case, the effective amount of currency linearly relates to the number of necessary transactions or transferable goods. More currency input may only impact the nominal price, which is classified as circulation inflation (Type I).
	
	5. The dominant demand amount of non-depreciable currency is the overall accumulation of reservation goods ever produced. The contradiction between constant face value of currency and depreciable goods leads to an intrinsic reservation inflation (Type II).

\section{Acknowledgment}
We appreciate the financial support from the National Natural Science Foundation of China (Grant No. 11505110), the Shanghai Pujiang Talent Program (Grant No. 16PJ1431900).

\end{document}